\documentstyle[12pt,twoside]{article}
\normalbaselines
\makeatletter
\@addtoreset{equation}{section}
\makeatother

\newcommand{\be}{\begin{equation}}
\newcommand{\ee}{\end{equation}}
\newcommand{\bea}{\begin{eqnarray}}
\newcommand{\eea}{\end{eqnarray}}

\title{Fock-Bargmann Representation of the Distorted
Heisenberg Algebra}

\bigskip \bigskip

\author{J. Oscar Rosas-Ortiz\thanks{e-mail:
                                 orosas@fis.cinvestav.mx}
\\ \\
      \small {\it Departamento de F\'\i sica,
                                     CINVESTAV-IPN}\\
      \small {\it A.P. 14-740, 07000 
                             M\'exico D.F., Mexico}}
              
     \bigskip \bigskip

\begin{document}

\maketitle
\date{}

\thispagestyle{empty}

\begin{abstract}

The dynamical algebra associated to a family of Isospectral 
Oscillator Hamiltonians, named {\it Distorted Heisenberg 
Algebra} because its dependence on a distortion parameter 
$w\geq 0$, has been recently studied. The connection of
this  algebra with the Hilbert space of entire analytic
functions of growth  (1/2, 2) is analyzed.

\end{abstract}

\bigskip\bigskip

{\footnotesize
\begin{description}
\item[Key-Words:] Coherent states, Heisenberg algebra
\item[PACS:] 03.65.-w, 11.30.Pb, 42.50.Dv 

\end{description}}

\vfill


\newpage

\baselineskip=18pt
\setcounter{page}{1}
\section{Introduction}

In 1980, Abraham and Moses found a general class of 
1-dimensional potentials isospectral to the oscillator one 
by means of the Gelfand-Levitan formalism \cite{1}. An 
elegant way to construc the same class, used by Mielnik 
\cite{2}, consists in the application of a variant of the 
standard factorization method. The connection between the 
Darboux transformation and this generalized factorization 
has been recently discussed \cite{3,4,5}. The Mielnik 
construction is suitable to easily identify a pair of 
annihilation and creation  operators for the isospectral 
oscillator Hamiltonians \{$A, A^{\dagger}$\}, which are 
adjoint to each other, although their  commutator is not the 
identity \cite{2}. Departing of the previous operators, a  
different pair, \{$A, B^{\dagger}$\}, was constructed such 
that the  commutator is the identity but $(A)^{\dagger} \neq 
B^{\dagger}$ \cite{6}. Recently, a third choice was done 
\cite{7}, where the annihilation  $C_w$ and creation 
$C_w^\dagger$ operators are adjoint to each other and  
commute to the identity on a subspace of the state space, 
imitating then  the behavior of the usual annihilation and 
creation operators for the  harmonic oscillator, {\it i.e.}, 
the Heisenberg-Weyl algebra. It turns  out that $C_w$ and 
$C_w^{\dagger}$ depend on a parameter $w \geq 0$. The  
apparence of this parameter is important because it leads to 
the  Heisenberg-Weyl algebra for some of its particular 
values. Moreover, the  coherent states constructed from 
these operators reach the standard form of the harmonic 
oscillator coherent states for such $w$-values. In the  
general case ($w$ arbitrary), these operators have an 
algebraic structure  very similar to the harmonic oscillator 
one. 

The Fock-Bargmann representation of the Heisenberg-Weyl 
algebra is widely used in physics and mathematics \cite{8}. 
The first application was made in quantum field theory, 
where the operators $\bar z$ and $\frac {\partial}{\partial 
\bar z}$ represent the creation and annihilation of bosons. 
Recently, this algebra has been considered also in the study 
of tensor bosons, arising in composite object theories, 
which according of their symmetry properties have been 
classified as symmetric or antisymmetric \cite{9}. 

The resemblance between the operator pair 
$\{C_w,C_w^{\dagger}\}$ and the Heisenberg-Weyl 
corresponding one $\{a,a^{\dagger}\}$, suggests the 
following question, which we will try to answer in this 
paper: which are the properties of the operators $C_w$ and 
$C_w^\dagger$ when they act on a space of entire analytic
functions? In particular, we will consider  just one 
of the two kinds of coherent states obtained by Fern\'andez 
{\it et al} \cite{7}, and we will show that the  
corresponding Fock-Bargmann space ${\cal F}_w$ contains 
entire analytic  functions of growth (1/2, 2). The 
realization of $C_w$ and $C_w^\dagger$  on ${\cal F}_w$ is a 
multiplication by $\bar z$ and a derivation with respect to 
$\bar z$ plus a new term dependent on the distortion 
parameter respectively. We discuss also the specific example 
with $w=1$, and we show that the usual harmonic oscillator 
case is recovered on ${\cal F}_1$. 

\section{Distorted Heisenberg algebra revisited}

Let us consider an infinite discrete set of orthonormal 
state vectors in the Hilbert space ${\cal H}, $\{ $\vert 
\theta_n \rangle, n=0,1,2,...$\}. This is a basis set in 
${\cal H}$ related to the standard harmonic oscillator basis 
\{ $\vert \psi_n \rangle, n=0,1,2,...$\} by \cite{2}:
\[
b \vert\theta_n\rangle =
\sqrt{n}\,\vert\psi_{n-1}\rangle,  \qquad
b^\dagger\vert\psi_n\rangle=\sqrt{n+1}\,\vert
\theta_{n+1}\rangle.
\]
The $\vert \theta_n \rangle$'s, $n=0, 1, 2,...$,
satisfy the following eigenvalues equation
\[
H_{\lambda} \vert \theta_n \rangle = E_n^{(\lambda)} 
\vert \theta_n\rangle
\]
where $ E_n^{(\lambda)}= E_n = n+1/2$, and $H_{\lambda} = 
b^{\dagger} b + 1/2$ is the Hamiltonian isospectral to the 
harmonic oscillator Hamiltonian $H=a^{\dagger} a + 1/2=a  
a^{\dagger} - 1/2$ obtained through the generalized  
factorization method of Mielnik \cite{2}. The explicit 
expression of  $H_{\lambda}$ in the coordinate 
representation is 
\[
H_{\lambda}= -\frac12 \frac{d^2}{dx^2} +
\frac{x^2}{2} -  \frac{d}{dx} \left[
\frac{e^{-x^2}}{\lambda + \int_0^x e^{-y^2} dy} 
\right] 
\]
Notice that the factorizing operators $b^{\dagger}$ and $b$ 
are not the raising and lowering operators for the basis 
\{$\vert \theta_n\rangle$\}. According to Fern\'andez {\it 
et al} \cite{7}, the raising and lowering operators similar 
to those of the harmonic oscillator, and leading to the 
distorted Heisenberg-Weyl algebra, are built by means of the 
condition: 
\[ 
[C,C^\dagger]=I \quad {\rm on} \quad {\cal H}_s 
\subset {\cal
H}.
\]
where ${\cal H}_s$ is a subspace of the Hilbert space ${\cal 
H}$. 

It turns out that these operators depend on a parameter $w 
\geq 0$ and take the form \cite{7}: 
\[ 
C_w=b^\dagger\frac{1}{N+1}\sqrt{\frac{N+w}{N+2}} ab, 
\qquad C_w^\dagger=b^\dagger
a^\dagger\frac{1}{N+1}\sqrt{\frac{N+w}{N+2}} b.
\]
Let's remark that since $b$, $b^{\dagger}$, $a$, 
$a^{\dagger}$ are first order differential operators, then 
$C_w$ and $C_w^{\dagger}$ are differential operators of 
order greater than three (indeed they are infinite order 
differential operators). Their action on the basis \{ $\vert 
\theta_n \rangle$\} is given by: 
\bea 
C_w\vert\theta_n\rangle & = &
(1-\delta_{n,0}-\delta_{n,1})\ \sqrt{n-2+w}\
\vert\theta_{n-1}\rangle,  \\ 
\nonumber
C_w^\dagger \vert\theta_n\rangle & = &
(1-\delta_{n,0})\
\sqrt{n-1+w}\ \vert\theta_{n+1}\rangle ,\\
\nonumber
I_w \vert\theta_n\rangle & = &
[1-\delta_{n,0} +
\delta_{n,1}(w -1)] \vert\theta_n\rangle,
\eea
where $I_w \equiv [C_w,C_w^\dagger]$. From this action, it 
is clear that for $w>0$ the set of operators \{$ C_w, 
C_w^{\dagger}, I_w$\} enables one to decompose the Hilbert 
space ${\cal H}$ as a direct sum of two invariant subspaces, 
${\cal H} = {\cal H}_0 \oplus {\cal H}_r$, where ${\cal 
H}_0$ is spanned by $\vert \theta_0\rangle$ and ${\cal H}_r$ 
by \{$\vert \theta_n \rangle, n \geq 1$\}. These two 
subspaces induce irreducible representations of $C_w$, 
$C_w^{\dagger}$ and $I_w$. The usual representation of the 
Heisenberg  algebra is recovered on ${\cal H}_r$ for $w=1$. 
On the other hand, for $w=0$ ${\cal H}$ decomposes (under 
the action of \{$C_0$, $C_0^{\dagger}$, $I_0$\}) as the 
direct sum of three invariant subspaces ${\cal H}={\cal H}_0 
\oplus {\cal H}_1 \oplus {\cal H}_s$, where ${\cal H}_1$ is 
generated by $\vert \theta_1 \rangle$ and ${\cal H}_s$ by \{ 
$\vert \theta_n \rangle, n \geq 2$ \}, and the usual 
Heisenberg algebra representation is recovered once again on 
${\cal H}_s$. In this paper we will work the case $w>0$, and 
when we consider a particular case we will analyze just the 
one with $w=1$. 

We can propose the new operator $N_w = C_w^{\dagger} C_w$, 
analogous to the standard number operator $N=a^\dagger a$, 
which relevant representation arises when we consider its 
action on vectors $\vert \psi \rangle \in {\cal H}_r$. The 
standard number operator representation is recovered on 
${\cal H}_r$ by taking $w=1$ in $N_w$ and relabeling the 
eigenstates of $H_{\lambda}$ as $\vert \phi_n \rangle \equiv 
\vert \theta_{n+1} \rangle $. However, the dependence of 
$N_w$ as a function of the Hamiltonian $H_{\lambda}$ is not 
obvious (in the general case $w>0$). A nonlinear dependence 
is probably in the context of the recent generalized Fock 
treatment \cite{10}, but it is quite involved to determine 
it precisely. This is due to the fact that the order of 
$C_w$ and $C_w^{\dagger}$ is in general infinity while the 
one of $H_\lambda$ is finite (two). 

The normalized ground state of $H_\lambda$, $\vert \theta_0 
\rangle$, can be introduced by means of the requirements 
$C_w \vert \theta_0 \rangle = C_w^{\dagger} \vert \theta_0 
\rangle = 0$. The solution to these  equations in the 
coordinate representation is given by \cite{6}: 
\[
\theta\sb0(x) \propto \frac{e\sp{-x\sp2/2}}
{\lambda+\int\sb0\sp x
e\sp{-y\sp2} dy}.
\]
where $\lambda \in {\bf R}$, $\vert \lambda \vert \,> 
\sqrt{\pi} /2 $. Hence $\vert \theta_0 \rangle $ is 
orthogonal to all $\vert \theta_n \rangle \in {\cal H}_r$ 
\cite{2}. Because $C_w \vert \theta_1\rangle=0$ (see 2.1), 
the eigenvalue $z=0$ of $C_w$ is doubly degenerated. As a 
consequence, the state $\vert \theta_0\rangle$ is 
disconnected of the space ${\cal H}_r$ and $\vert \theta_1 
\rangle$ is the state playing the role of the extremal state 
for the distorted Heisenberg-Weyl algebra. Thus, the 
operator $C_w^{\dagger}$ can be used to construct the basis 
of the state space ${\cal H}_r$ from its repeated action on 
$\vert \theta_1\rangle$: 
\be
\vert\theta_n\rangle=
\sqrt{\frac{\Gamma(w)}{\Gamma(w+n-1)}}
(C_w^\dagger)^{n-1}\vert\theta_1\rangle, \quad \quad 
w \neq 0.
\ee
A similar situation has been recently reported by Spiridonov 
for systems and creation and annihilation operators 
different to the ones used in this paper (see \cite{4}, 
section VII). Spiridonov also looks for the coherent states 
as eigenstates of the corresponding annihilation operator, 
and they could be found through the solution of some third 
order differential equations ((7.10) and (7.13) of 
\cite{4}). Unfortunately, be means of such a method one 
cannot find easily and directly those coherent states;
however, it can be done for the isospectral Hamiltonians we
are dealing with \cite{6,7}. This is one of the advantages
of the coherent states we will present in the next section.

As a final point of this section, the resolution of the 
identity in terms of the basis \{$ \vert \theta_n\rangle, 
n\geq 0$\} is the standard one 
\[
I=\sum_{n=0}^{\infty} \vert \theta_n \rangle \langle 
\theta_n \vert.
\]
It let us to expand any state vector $\vert
h\rangle \in {\cal H}$ as:
\[
\vert h \rangle = \sum_{n=0}^{\infty} a_n \vert 
\theta_n \rangle; \quad \quad a_n \equiv \langle 
\theta_n \vert h\rangle.
\]
\section{Distorted Heisenberg algebra Coherent States}

Recently, a family of coherent states has been
constructed as eigenstates of the annihilation operator,
$C_w \vert z,w \rangle = z\vert z,w\rangle$; they are
called here $w$-{\it  coherent states}. Their explicit
form in terms of $\vert \theta_n\rangle$ is given by
\cite{7}:
\be
\vert z,w\rangle =\sqrt{\frac{\Gamma(w)}{
{}_1F_1(1,w;r^2)}}\
\sum_{n=0}^{\infty} \frac{z^n}{\sqrt{\Gamma(w+n)}}
\ \vert \theta_{n+1}\rangle, \quad \quad z=r 
e^{i\varphi}
\ee
where ${}_1F_1(a,b;x)$ is a hypergeometric function.
The ket $\vert \theta_0  \rangle$, by construction, is
also a coherent state. The set \{ $\vert \theta_0
\rangle, \vert z, w \rangle$ \} is complete in ${\cal
H}$,  with an identity resolution
\be
I=\vert
\theta_{0}\rangle\langle\theta_{0}\vert + \int
\vert z,w \rangle\langle z,w\vert  \, d\mu(z,w)
\ee
where $d\mu(z,w)$ is the measure written as:
 \be
d\mu(r,w)= \frac{{}_1F_1(1,w;r^2)}{\pi\Gamma(w)}\ 
e^{-r^2}\
r^{2(w -1)} \, r \, dr\, d\varphi.
\ee

As usual, this new basis set is not orthogonal because
the inner product between two different coherent states
\be
\langle z,w \vert z',w \rangle = \frac{_1F_1(1,w; \bar z
z')}{\sqrt{_1F_1(1,w;r^2)  \ {}_1F_1(1,w;r'^2)}} 
\ee 
is in general, different from zero. Any vector state
$\vert h\rangle \in {\cal H}$ is now expanded as
\be
\vert h\rangle=h_0 \vert \theta_0\rangle+\int 
{\tilde h} (z,\bar
z ,w)
\vert z,w\rangle d\mu(z,w),
\ee
where $h_0\equiv \langle\theta_0\vert h\rangle$, 
\be
{\tilde h} (z,\bar z, w)\equiv \langle z,w\vert h
\rangle=
\frac{h_w(\bar z)}{\sqrt{{}_1F_1(1,w;r^2)}},
\ee 
and
\be
h_w(\bar z)\equiv \sum_{n=0}^{\infty} \sqrt{\frac
{\Gamma(w)}{\Gamma(w+n)}} \bar z^n \langle\theta_{n+1}
\vert h\rangle.
\ee 
In this representation, the inner product between
$\vert f \rangle \in {\cal H}$ and $\vert g \rangle \in
{\cal H}$ is:
\bea
\langle f \vert g \rangle & = & \bar f_0 g_0
+ \int \frac{{\bar f_w(\bar z)} g_w(\bar z)}
{{}_1F_1(1,w,r^2)} d\mu(\bar z,w),\\
\nonumber
& = & \bar f_0 g_0 +\int 
{\bar f_w(\bar z)} g_w(\bar z) d\sigma_w (\bar z),
\eea
where
\be
d\sigma_w(z)= \frac{e^{-r^2}\ r^{2(w
-1)}}{\pi\Gamma(w)} \, r \, dr\, d\varphi.
\ee 
In the special case, $f=g$ we obtain:
\be
\vert \vert \, \vert f \rangle \, \vert \vert^2 = 
\vert f_0\vert^2 + \int \vert f_w(\bar z) 
\vert^2 d \sigma_w(z) \, \geq 0
\ee
\section{The Hilbert space of
Entire Functions.}
\subsection{Basic properties of ${\cal F}_w$.}

The goal of this section is to characterize any vector
state by an entire analytic function. Such a
realization of the Hilbert space ${\cal H}$ is called
the {\it Fock-Bargmann representation} \cite{11,12}, and
it can be obtained through any coherent state
system. Here, we will consider the coherent states 
$\vert z,w\rangle$ and the Hilbert space ${\cal H}_r 
\subset {\cal H}$.

Let us introduce the Hilbert space
${\cal F}_w$, which elements are all functions of the
form (3.7). For each $w$, the inner product is
defined by:
\be
(f,g)_w \equiv \int {\bar f_w(\bar z)} g_w(\bar z)
d\sigma_w (\bar z)
\ee
where the integral is extended over all ${\bf C}$. In
particular, for $f=g$ we require (see 3.10)
\be
(f,f)_w = \int \vert f_w(\bar z) \vert^2 d\sigma_w (\bar
z)<\infty.
\ee
It can be easily seen that $f_w(z)$ is an entire analytic
function (see 3.7). We want to study in more detail the 
properties of $f_w(z)$. With this aim, let us analyze
equation (3.6). Because $\vert z,w\rangle$ is normalized,
the  Schwarz inequality gives: $\vert \langle z,w \vert f
\rangle \vert \leq \vert \vert \, \vert f \rangle \,
\vert \vert $, $\vert f \rangle \in {\cal H}_r$, and one
obtains:
\be
\vert f_w(z)\vert \leq \sqrt{ {}_1F_1 (1,w,r^2)} \quad
\vert \vert  \, \vert f \rangle \, \vert \vert.
\ee
Hence, the behavior of $\vert f_w(z)\vert$ at infinity
is the same as the one of $\sqrt{ {}_1F_1 (1,w,r^2)}$. As
it has been shown, the generalized hypergeometric function
\bea
&& {}_pF_q (a_1,..., a_p, b_1, ..., b_q, z^s ) 
\equiv {}_pF_q (a_i, b_j, z^s)\\
\nonumber
& = & \frac{ \Gamma(b_1) ... \Gamma(b_q)}
{\Gamma(a_1) ... \Gamma(a_p)}
\sum_{n=0}^{\infty} \frac{\Gamma(a_1+n) ... 
\Gamma(a_p+n)}{ \Gamma(b_1 + n) ... \Gamma(b_q + n)} 
\, \frac{z^n}{n!}
\eea
is an entire analytic function (exponential type) of
order $\rho_s$ and type $\sigma_s$ \cite{6}, {\it
i.e.}, growth ($\sigma_s, \rho_s$). The relation between
$\rho_s$ and $\sigma_s$ is given by $\sigma_s = s/
\rho_s = 1 + q - p$. In the particular case of ${}_1F_1
(1,w,r^2)$ we find $\rho_s=2$ and $\sigma_s=1$. Then,
because:
\be
\vert \sqrt{ {}_pF_q(a_i,b_j;z^s)} \vert = \sqrt
{ \vert {}_pF_q (a_i,b_j, z^s) \vert } \leq \exp( 
\frac{\sigma_s}{2} r^{\rho_s})
\ee
one obtains that $\sqrt{ {}_1F_1 (1,w,r^2)}$ has
$\rho=2$ and $\sigma=1/2$, {\it i.e.}, growth (1/2,2).
This behavior at infinity is equal to the
corresponding one for the harmonic oscillator case,
which shows that the $w$ coherent states are
``good" to generate the Fock-Bargmann representation
of ${\cal H}_r$. Thus, we have realized the Hilbert
space ${\cal H}_r$ as a space ${\cal F}_w$ of entire
analytic functions $f_w(z)$ of the form (3.7) and 
satisfying condition (4.2).

The orthonormal basis \{$\vert \theta_n \rangle, n \geq
1 $\} in ${\cal F}_w$ has a simple form
\be
\theta_{n} (\bar z) \equiv \sqrt{ {}_1F_1 (1, w,r^2 )} \
\ \langle z,w  \vert \theta_n \rangle ={\bar z}^{(n-1)}
\sqrt{\frac {\Gamma (w)}{\Gamma (w+n-1)}}.
\ee
This is the simplest orthonormal set of vectors in
${\cal F}_w$. The functions $h_w(\bar z)$ are then
written as:
\be
h_w(\bar z) = \sum_{n=0}^{\infty} A_{n+1} \theta_{n+1} 
(\bar z); \quad A_{n+1} \equiv \langle \theta_{n+1} 
\vert h \rangle,
\ee
and the corresponding representation of the coherent
state $\vert \alpha,w\rangle, \quad \alpha \in {\bf C}$ 
is:
\bea
\alpha_w(\bar z) &=& \sqrt{{}_1F_1(1,w,r^2)} \
\ \langle z,w \vert \alpha, w \rangle\\
\nonumber
&=& \frac {{}_1F_1 (1,w,\bar z \alpha )}{\sqrt {{}_1F_1
(1,w,\vert \alpha \vert^2)}}.
\eea
Notice that (4.6) and (4.8) reproduce the usual
harmonic oscillator case when we take $w=1$.

\subsection{Principal vectors ${\bf e}_a$, and the
reproducing kernel}

Equation (4.3) is quite typical for a Hilbert space of
analytic functions, and the term $\sqrt {{}_1F_1
(1,w,r^2)}$ arises due to the set of coherent states taken
into  account. In this section we will derive some results
independent of the function ${}_1F_1 (1,w,r^2)$ and valid
for all possible spaces ${\cal F}$.

Let us consider the elements ${\bf
e}_a$ of ${\cal F}_w$ (for every $a \in {\bf C}$) given by:
\be
{\bf e}_a (z) \equiv \sum_{n=0}^{\infty} \theta_{n+1} 
(z) {\bar \theta_{n+1} (a)} = {}_1F_1 (1,w,z \bar a).
\ee
Using these vectors, called the {\it principal vectors}
of ${\cal F}_w$, we can introduce a bounded
linear functional:
\be
f_w (a)= ({\bf e}_a,f)_w.
\ee
The integral form reads:
\be
f_w (a)= \int {}_1F_1 (1,w,a \bar z) f_w (z) d
\sigma_w(z).
\ee
Then, ${}_1F_1 (1,w,a \bar z)$ is the reproducing kernel
for ${\cal F}_w$. Conversely, (4.10) implies
(4.3) with $\vert \vert {\bf e}_a \vert \vert = {}_1F_1
(1,w,\vert a \vert^2)$.

The connection between the reproducing kernel and the
coherent states as represented in the ${\cal F}_w$
space is given by (4.8). This could be seen
also by expressing (4.11) in terms of the $w$ coherent
states $\vert z,w \rangle \in {\cal H}_r$ because:
\bea
\nonumber
f_w (\bar a) &=& \sqrt{ {}_1F_1 (1,w,\vert a \vert^2)} \langle
a,w \vert f \rangle\\
\nonumber
&=& \int \frac{ {}_1F_1 (1,w, \bar a z)}{\sqrt{
{}_1F_1(1,w,r^2)}} \langle z,w \vert f \rangle
d \mu (z,w)\\
\nonumber
&=& \int {}_1F_1 (1,w,z \bar a) f_w (\bar z) d
\sigma_w(z).
\eea
\noindent
>From this, we get the proportionality between the
reproducing kernel and the inner product of two $w$
coherent states $\vert a, w \rangle$ and $\vert z, 
w \rangle$ both in ${\cal H}_r$.

Finally, the set of vectors ${\bf e}_a$ is complete, {\it
i.e.} their finite linear combinations are dense in
${\cal F}_w$, because the only vector orthogonal to all of
them is $f=0$, as follows immediately from (4.10).

\subsection{Realization of the distorted Heisenberg
Algebra in ${\cal F}_w$}

Let us represent now the generators of the distorted
algebra in ${\cal F}_w$, which is a subspace of the space of entire 
functions of growth (1/2,2). At this point, it is convenient to
introduce the unnormalized $w$ coherent states $\vert z,w
\rangle_e = \sqrt{ {}_1F_1(1,w,r^2)} \vert z,w \rangle$.
In terms of these the function $h_w(\bar z)$ can be
written as:
\[
h_w(\bar z)={}_e\langle z,w\vert h\rangle.
\]
The action of $C_w^{\dagger}$ on $h_w(\bar z)$ arises
from the inner product of $\vert h \rangle$ with the
adjoint of $ C_w\vert z,w\rangle_e= z \vert z, w
\rangle_e$:
\be
C_w^\dagger h_w(\bar z) \equiv {}_e \langle z,w\vert 
C_w^\dagger \vert h\rangle = {\bar z} h_w(\bar z).
\ee
The action of $C_w$ on $h_w(\bar z)$ is as well
obtained from the inner product of $\vert h
\rangle$ with the adjoint of
\bea
C_w^\dagger \vert z,w\rangle_e &=& \sum_{m=1}^{\infty} 
{\sqrt \frac{\Gamma(w)}{\Gamma(w+m)}} z^{m-1}m \vert 
\theta_{m+1}\rangle+\\
\nonumber
& & \frac{(w-1)}{z} \sum_{n=0}^{\infty} {\sqrt 
\frac{\Gamma(w)}{\Gamma(w+n)}} z^n \vert 
\theta_{n+1}\rangle-\frac{(w-1)}{z}\vert 
\theta_1\rangle.
\eea
We get then:
\be
C_w h_w(\bar z)=\frac{\partial}{\partial \bar z} 
h_w(\bar z)+ \frac{(w-1)}{\bar z} \left( h_w(\bar z)-h_1
\right),
\ee
where $h_1\equiv \langle \theta_1\vert h\rangle$, $\bar
z\neq 0$. Finally, the action of $I_w$ on $h(\bar z)$ is
given by:
\be 
I_w h_w(\bar z) = (w-1) h_1 + h_w (\bar z).
\ee

At this point we see the great resemblance, almost
equality, of equations (4.12, 4.14, 4.15), with the
corresponding ones for the harmonic oscillator case. The
difference rests on the  dependence on $w>0$. When $w=1$,
however, (4.12, 4.14, 4.15) are the same as those for the
usual harmonic oscillator in the Segal-Bargmann space.
There is an isomorphism then between ${\cal F}_w$ and
${\cal H}_r$ for all $w>0$. Notice that it is also 
possible to find an isomorphism between ${\cal F}_w$ and
a subspace of ${\cal H}$ for $w=0$, but in this case it
will be ${\cal H}_s$.

We would like to show, finally,  that the coherent states
$\alpha_w(z) \in {\cal F}_w$, expressed in (4.8) in terms
of  $\vert z,w \rangle \in {\cal H}_r$, can be obtained
through a  standard procedure \cite{6,7} as
eigenfunctions of $C_w$ in ${\cal F}_w$:
\be
C_w \alpha_w (\bar z)= \alpha \ \alpha_w(\bar z), \ \
\alpha
\in {\bf C}.
\ee
Let's take
\be
\alpha_w (\bar z)= \sum_{n=1}^{\infty} a_n \theta_n(\bar
z).
\ee
Using (4.14), with $h_w(\bar z) = \alpha_w (\bar
z)$, and (4.16) one obtains the coefficients $a_n$. In the
case $w \neq 0$ we get
\be
a_{n+1}= \alpha^n \sqrt{\frac{\Gamma(w)}{\Gamma(w+n)}} 
a_1, \
\ n=1,2,...
\ee
Then, by imposing the normalization, with $a_1 \geq 0$, we
obtain
\[
\alpha_w(\bar z)= \frac { {}_1F_1 (1,w, \alpha \bar z)}{
\sqrt{ {}_1F_1 (1,w, \vert \alpha \vert^2)}}
\]
which, obviously coincides with (4.8). Notice that the
case $w=1$ is straightforwardly obtained of the previous
formula:
\[
\alpha_1 (\bar z)= {\rm e}^{(-\frac12 \vert \alpha
\vert^2 + \alpha {\bar z})}
\]
This is the usual representation for the harmonic
oscillator coherent state in the Segal-Bargmann space.

All these facts show that $C_w$ and $C_w^\dagger$, adjoint
to each other in ${\cal H}_r$, possess the same 
properties on ${\cal F}_w$. Hence, the eigenstates of
$C_w$ in ${\cal H}_r$ correspond to eigenfunctions of
$C_w$ in ${\cal F}_w$ and vice versa.

\section*{Remarks}
The $w$ coherent states $\vert z, w\rangle$ and the
operators $C_w, \ C_w^\dagger$ and $I_w$, admit a
simplest  representation in the space of analytic
functions ${\cal F}_w$ compared with the representation
in ${\cal H}_r$. The  usual harmonic oscillator
representation is achieved when $w=0$ or $w=1$. The fact
that $C_w$ and $C_w^\dagger$ are adjoint to each other on
${\cal H}_r$ is
preserved on ${\cal F}_w$ because the isomorphism 
between ${\cal F}_w$ and ${\cal H}_r$. This
representation of the  distorted Heisenberg algebra on
${\cal F}_w$ as far as we know, has not  been explored
previously. It could be important in the analysis of the 
appropriate displacement operator $D_w$ to perform the
Perelomov  construction of the coherent states for the
isospectral oscillator  Hamiltonians $H_{\lambda}$.

\section*{Acknowledgements}

The author acknowledges CONACyT (M\'exico) for financial 
support, and to Dr. David J. Fern\'andez C. for critical 
observations and remarks. The suggestions of the referees of 
this paper are also acknowledged.

\vfill\eject

\end{document}